\documentclass[preprint,aps,superscriptaddress]{revtex4}

\usepackage{graphicx}
\usepackage{subfigure}
\usepackage{epsfig}
\usepackage{dcolumn}
\usepackage{bm}
\usepackage{ulem}
\usepackage{color}

\usepackage{epstopdf}

\def\be{\begin{equation}}       \def\ee{\end{equation}}
\def\bea{\begin{eqnarray}}      \def\eea{\end{eqnarray}}
\def\ba{\begin{array} }
\def\ea{\end{array} }
\def\bnum{\begin{enumerate} }
\def\enum{\end{enumerate}}

\def\=>{\Rightarrow}
\def\>{\rightarrow}

\def\eye2{Fathbb{I}}

\renewcommand{\>}{\rangle}

\begin{document}

\title{
Fully gapped topological surface states in Bi$_2$Se$_3$ films induced by a $d$-wave high-temperature superconductor}

\author{Eryin Wang}
\altaffiliation{These authors contribute equally to this work.}
\affiliation{State Key Laboratory of Low Dimensional Quantum Physics and Department of Physics, Tsinghua University, Beijing 100084, China}
\affiliation{Advanced Light Source, Lawrence Berkeley National Laboratory, Berkeley, CA 94720, USA}

\author{Hao Ding}
\altaffiliation{These authors contribute equally to this work.}
\affiliation{State Key Laboratory of Low Dimensional Quantum Physics and Department of Physics, Tsinghua University, Beijing 100084, China}

\author{Alexei V. Fedorov}
\affiliation{Advanced Light Source, Lawrence Berkeley National Laboratory, Berkeley, CA 94720, USA}

\author{Wei Yao}
\affiliation{State Key Laboratory of Low Dimensional Quantum Physics and Department of Physics, Tsinghua University, Beijing 100084, China}

\author{Zhi Li}
\affiliation{State Key Laboratory of Low Dimensional Quantum Physics and Department of Physics, Tsinghua University, Beijing 100084, China}

\author{Yan-Feng Lv}
\affiliation{State Key Laboratory of Low Dimensional Quantum Physics and Department of Physics, Tsinghua University, Beijing 100084, China}

\author{Kun Zhao}
\affiliation{State Key Laboratory of Low Dimensional Quantum Physics and Department of Physics, Tsinghua University, Beijing 100084, China}

\author{Li-Guo Zhang}
\affiliation{State Key Laboratory of Low Dimensional Quantum Physics and Department of Physics, Tsinghua University, Beijing 100084, China}

\author{Zhijun Xu}
\affiliation{Condensed Matter Physics and Materials Science Department, Brookhaven National Laboratory, Upton, New York 11973, USA}

\author{John Schneeloch}
\affiliation{Condensed Matter Physics and Materials Science Department, Brookhaven National Laboratory, Upton, New York 11973, USA}

\author{Ruidan Zhong}
\affiliation{Condensed Matter Physics and Materials Science Department, Brookhaven National Laboratory, Upton, New York 11973, USA}

\author{Shuai-Hua Ji}
\affiliation{State Key Laboratory of Low Dimensional Quantum Physics and Department of Physics, Tsinghua University, Beijing 100084, China}
\author{Lili Wang}
\affiliation{Institute of Physics, Chinese Academy of Sciences, Beijing 100190, China}
\author{Ke He}
\affiliation{Institute of Physics, Chinese Academy of Sciences, Beijing 100190, China}
\author{Xucun Ma}
\affiliation{Institute of Physics, Chinese Academy of Sciences, Beijing 100190, China}

\author{Genda Gu}
\affiliation{Condensed Matter Physics and Materials Science Department, Brookhaven National Laboratory, Upton, New York 11973, USA}
\author{Hong Yao}
\affiliation{Institute for Advanced Study, Tsinghua University, Beijing 100084, China}
\author{Qi-Kun Xue}
\affiliation{State Key Laboratory of Low Dimensional Quantum Physics and Department of Physics, Tsinghua University, Beijing 100084, China}
\author{Xi Chen}
\altaffiliation{Correspondence should be sent to xc@mail.tsinghua.edu.cn and syzhou@mail.tsinghua.edu.cn}
\affiliation{State Key Laboratory of Low Dimensional Quantum Physics and Department of Physics, Tsinghua University, Beijing 100084, China}
\author{Shuyun Zhou}
\altaffiliation{Correspondence should be sent to xc@mail.tsinghua.edu.cn and syzhou@mail.tsinghua.edu.cn}
\affiliation{State Key Laboratory of Low Dimensional Quantum Physics and Department of Physics, Tsinghua University, Beijing 100084, China}

\date{\today}

\begin{abstract}

{\bf Topological insulators are a new class of materials \cite{HasanRMP, QiZRMP}, that exhibit robust gapless surface states protected by time-reversal symmetry \cite{AliScattering,XueScattering}. The interplay between such symmetry-protected topological surface states and symmetry-broken states (e.g. superconductivity) provides a platform for exploring novel quantum phenomena and new functionalities, such as 1D chiral or helical gapless Majorana fermions \cite{KanePRB10}, and Majorana zero modes \cite{FKProximity} which may find application in fault-tolerant quantum computation \cite{Kitaev,RMPQCP}. Inducing superconductivity on topological surface states is a prerequisite for their experimental realization \cite{HasanRMP, QiZRMP}. Here by growing high quality topological insulator Bi$_2$Se$_3$ films on a $d$-wave superconductor Bi$_2$Sr$_2$CaCu$_2$O$_{8+\delta}$ using molecular beam epitaxy, we are able to induce high temperature superconductivity on the surface states of Bi$_2$Se$_3$ films with a large pairing gap up to 15 meV. Interestingly, distinct from the $d$-wave pairing of Bi$_2$Sr$_2$CaCu$_2$O$_{8+\delta}$, the proximity-induced gap on the surface states is nearly isotropic and consistent with predominant $s$-wave pairing as revealed by angle-resolved photoemission spectroscopy. Our work could provide a critical step toward the realization of the long sought-after Majorana zero modes.}

\end{abstract}

\maketitle

The search for exotic quantum phenomena and novel functionalities has been among the most tremendous driving forces for the fields of condensed matter physics and materials science. Majorana zero modes, i.e. Majorana fermions which are their own antiparticles  and occur at exactly zero energy, are particularly fascinating not only because of their intriguing physics obeying robust non-Abelian statistics, but also due to their potential application as building blocks for topological quantum computer \cite{Kitaev,RMPQCP}. Although significant progresses have been made recently in one dimensional semiconductor quantum wires coupled with conventional superconductors \cite{MajoranaSci,DasNaturePhys,Josephson,XuNanoLett}, decisive evidences of Majorana zero modes have been lacking and many puzzles remain \cite{FranzReview}. Topological insulators (TIs), whose hallmark is time-reversal symmetry protected surface states (SS), may offer less restrictive experimental conditions for realizing Majorana zero modes \cite{HasanRMP, QiZRMP}. Theoretically, Majorana zero modes are predicted to occur in vortex cores of three dimensional TIs when they are in close proximity to conventional s-wave superconductors \cite{FKProximity}; however, identifying them experimentally has been challenging mainly because of the small pairing gap $\Delta\sim 1$ meV induced by typical conventional superconductors \cite{QianDBi2Se3} and the extremely small energy splitting $\sim \Delta^2/\epsilon_F\sim 10^{-3}$ meV between Majorana zero modes and other low-lying vortex core bound states \cite{degennes-book}, where $\epsilon_F\sim 400$ meV is the typical Fermi energy of SS \cite{QianDBi2Se3}. Heterostructure between a topological insulator and a cuprate high temperature superconductor (Fig.~1a), which has an order of magnitude enhancement in $T_c$ and $\Delta$ compared with typical conventional $s$-wave superconductors, may offer a more feasible approach \cite{Nagaosa, AdvCuprate}.

To harbor Majorana zero modes with robust non-Abelian statistics, nodeless superconductivity needs to be induced on the SS of TIs. Although previous tunneling junction experiments have suggested proximity-induced gap of 10 meV at the interface between exfoliated bulk Bi$_2$Se$_3$ and Bi$_2$Sr$_2$CaCu$_2$O$_{8+\delta}$ (Bi2212) samples \cite{BurchNatCom}, two critical questions remain unanswered for the proximity-induced superconductivity: whether the superconducting gap is induced on the SS or on the bulk states, and what is the pairing symmetry. To address these issues and achieve high $T_c$ superconductivity in TI, we prepare high quality Bi$_2$Se$_3$ thin films directly on freshly cleaved Bi2212 surface by molecular beam epitaxy (MBE). MBE growth can provide high quality TI films and controllable interface. More importantly, MBE films are suitable for ultra-high vacuum (UHV) measurements, such as scanning tunneling microscopy (STM) and angle-resolved photoemission spectroscopy (ARPES) \cite{DamascelliReview}. The latter can be conveniently used to determine whether the gap is on the topological SS or on the bulk bands, measure the gap size and reveal the pairing symmetry. Our ARPES measurements reveal a proximity-induced gap up to 15 meV, and this apparent gap is located only on the topological SS with predominant $s$-wave pairing. One possible explanation for this unusual proximity effect is the combined consequence of disorder and the relative lattice orientation between cuprate and TI film. Our work not only provides new opportunities for investigating the intriguing coupling between a TI thin film and a $d$-wave superconductor, but also greatly expands the experimental range for realizing robust Majorana zero modes and achieving novel functionalities.

\begin{figure*}
\includegraphics[width=16.8 cm] {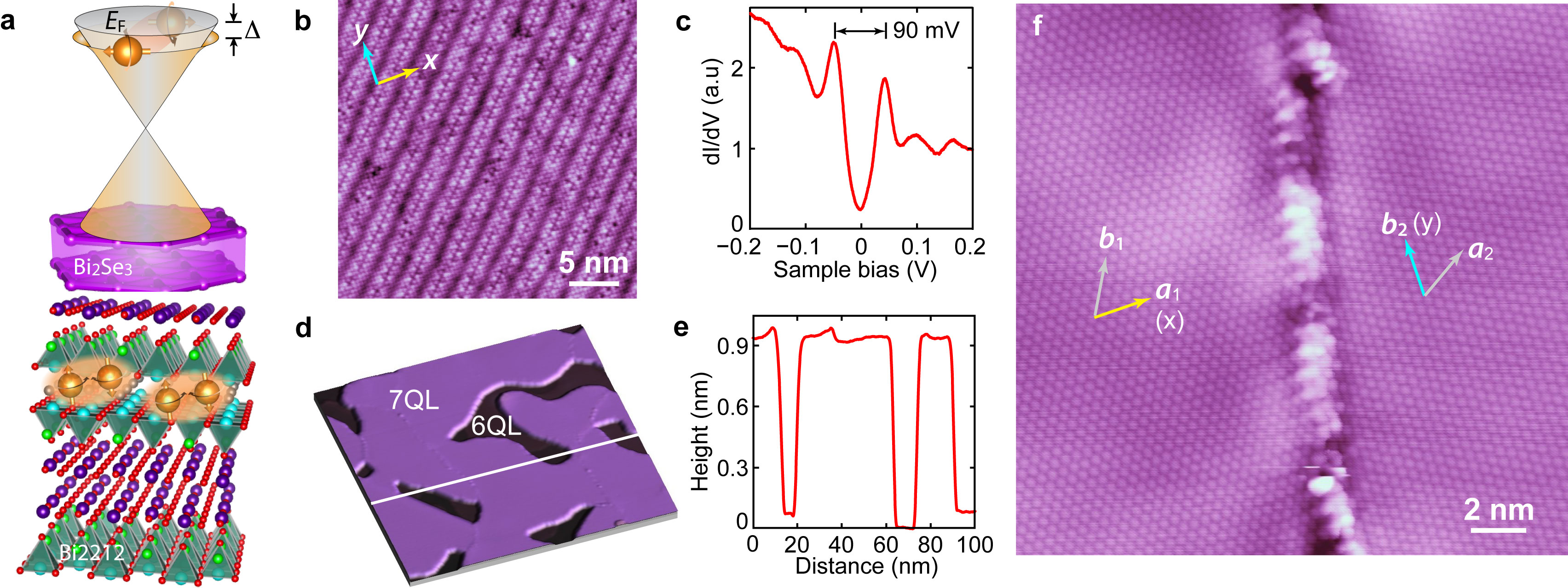}
\label{Figure 1}
\caption{{\bf Schematic and characterization of Bi$_2$Se$_3$/Bi2212 heterostructure.} (a) Schematic of Bi$_2$Se$_3$/Bi2212 heterostructure and proximity-induced gap on the surface states. (b) STM topography (sample bias voltage 0.3 V, tunneling current 0.03 nA) of BiO surface. (c) Typical dI/dV spectrum showing a superconducting gap of 45 meV at 4.2 K on the surface of Bi2212. (d) STM topography (100 nm $\times$ 100 nm, 1.0 V, 0.03 nA) of 7 QL Bi$_2$Se$_3$ films grown on Bi2212. (e) Line profile along the white line in (d). (f) Atomic-resolved STM topography (0.4 mV, 0.07 nA) shows two predominant crystal orientations of Bi$_2$Se$_3$.}
\end{figure*}

Figure 1a shows the schematic of the Bi$_2$Se$_3$/Bi2212 heterostructure. The STM topography (Fig.1b) and scanning tunnelling  spectroscopy (Fig.1c) confirm the high surface quality and superconducting property of the cleaved Bi2212 sample. The growth of Bi$_2$Se$_3$ thin films on Bi2212 is exceptionally challenging because of two reasons: (1) the low surface binding energy of Bi2212 prefers island growth mode rather than layer-by-layer mode; (2) oxygen dopants in Bi2212 can easily oxidize the films grown on it when heated above 200 $^\circ$C, a typical temperature for growing Bi$_2$Se$_3$. By carefully controlling the growth kinetics, we have succeeded in growing high quality Bi$_2$Se$_3$/Bi2212 heterostructures in the layer-by-layer growth mode so that high resolution STM and ARPES measurements can be carried out.  Figs.1d-f show the high quality of the 7 QL thick Bi$_2$Se$_3$ film on Bi2212. Although Bi$_2$Se$_3$ and Bi2212 have very different (hexagonal versus tetragonal) crystal structures, the Bi$_2$Se$_3$ films grown on Bi2212 show only two predominant crystal orientations with Se-Se bonds parallel to x and y (anti-nodal, nearest Bi-Bi bond) directions of Bi2212, respectively (see Fig.~1b and Fig.~1f and supplementary information).

\begin{figure*}
\includegraphics[width=13.6 cm] {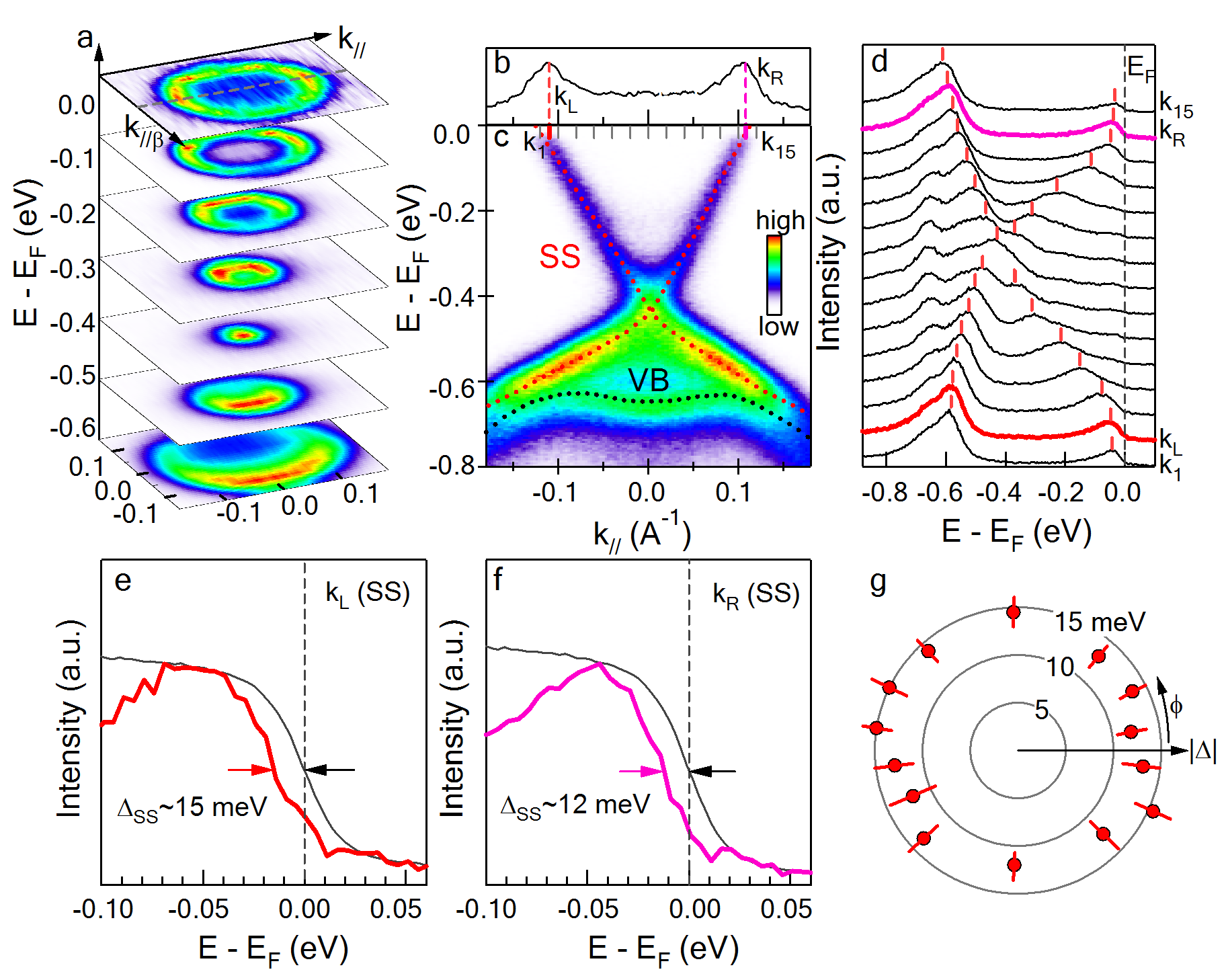}
\label{Figure 2}
\caption{{\bf ARPES data measured at 20 K with 50 eV photon energy reveal the nearly isotropic gap on the topological surface states.} (a) Constant energy maps. (b) MDC integrated from $-20$ to 20 meV. (c) ARPES data. Dotted lines are guides for the dispersions of SS and VB. (d) EDCs at momentum positions marked in (c). Red tick marks are guides for the peak positions of the SS. (e,f) Zoom-in of EDCs at $k_L$ (e) and $k_R$ (f). Black lines are reference spectra for E$_F$.  (g) Azimuth ($\phi$) dependence of the gap size (radial axis) along the Fermi surface.}
\end{figure*}

The existence of topological SS in Bi$_2$Se$_3$/Bi2212 heterostructure can be revealed in the ARPES data measured at 50 eV photon energy shown in Fig.2. Near the Fermi energy ($E_F$), the intensity map of the Dirac cone shows a circular shape, suggesting that the electronic structure is rather isotropic. The electron pocket decreases in size with binding energy and merges into a single point at -0.4 eV (Fig.2a). Figure 2c shows a cut through the $\Gamma$ point. Here the conical dispersion is observed from E$_F$ to $-0.6$ eV, where the bulk valence bands (VBs) start to arise. The electronic structure is in good agreement with the conical dispersion reported on the SS in both bulk and thin film Bi$_2$Se$_3$ samples \cite{ChenYL,HasanBi2Se3,XueBi2Se3,QianDBi2Se3}. The shift of chemical potential to 0.4 eV above the Dirac point is possibly attributed to charge transfer from the substrate and to defects in the Bi$_2$Se$_3$ film. In addition to SS and VBs, bulk conduction bands, shown as quantum well-like states (QWSs) near the Fermi energy, are also expected in thin film Bi$_2$Se$_3$ samples \cite{XueBi2Se3,QianDBi2Se3}. However, the SS and QWSs have different dipole matrix elements \cite{PanPRL}, which makes it possible to separate their contributions by choosing an appropriate photon energy (see supplementary information for detailed photon energy dependence). At 50 eV photon energy, the QWSs are strongly suppressed, thus allowing us to resolve the SS without being affected by the bulk bands.

\begin{figure*}
\includegraphics[width=14.2 cm] {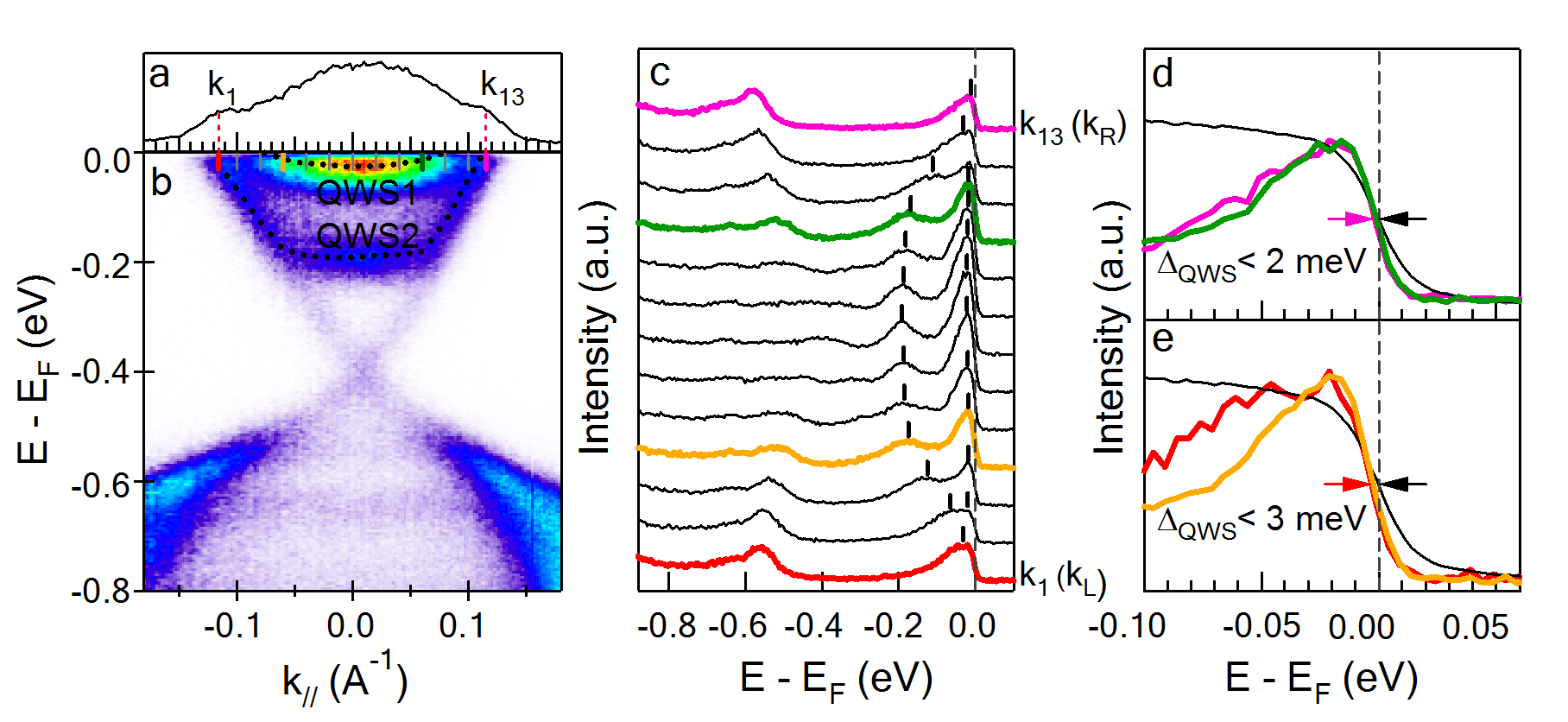}
\label{Fig3}
\caption{{\bf ARPES data measured at 20 K with 30 eV photon energy show a much smaller gap on the QWSs.} (a) MDC integrated from $-20$ to 20 meV. (b) ARPES data. (c) EDCs at momentum positions marked in panel (b). Black tick marks are guides for the peak positions of the QWSs. (d,e) Zoom-in of EDCs at Fermi momenta (colored curves in (c)) near $E_F$. Here the gap on the QWSs is reduced to less than 3 meV.}
\end{figure*}

We analyze the energy distribution curves (EDCs) taken at photon energy of 50 eV (Fig.2d) to detect the proximity-induced gap on the SS of Bi$_2$Se$_3$. The linear dispersions of the SS can be observed in the EDCs, with peaks approaching the Fermi energy at two Fermi momenta $k_L$ and $k_R$. We further zoom-in the EDCs at these two momentum positions (Figs.2e-f). A comparison of the EDCs with a reference spectrum taken from a clean metal surface shows clearly that there is a suppression of intensity near $E_F$, i.e. a gap is open. Following the standard ARPES procedure to extract the gap by the shift of leading edge position \cite{ShenLEG}, we obtain the gap size of 15 meV on the topologically protected SS. We note that although such large gap can be observed only on films with extremely high quality (sharper ARPES peaks), a smaller gap of 6-8 meV is routinely observed in regular samples measured so far, which is still an order of magnitude larger than that induced by a conventional $s$-wave superconductor \cite{QianDBi2Se3}. Surprisingly, despite the $d$-wave pairing of Bi2212 which is strongly anisotropic with four nodes of zero gap \cite{ShenLEG}, the induced gap on the SS is rather isotropic (Fig.2g) and consistent with $s$-wave pairing.

We examine the data taken at 20 K with photon energy of 30 eV where there is strong intensity contribution from QWSs to check if a gap is induced on the bulk conduction bands. The strong intensity contribution from QWSs is evident in Figs.3a-3c. Here the lower QWS (labeled as QWS2) is clearly resolved to overlap with the SS between E$_F$ and $-200$ meV with a nearly flat band bottom at $-200$ meV. The inner QWS (labeled as QWS1) shows up as a smaller pocket near E$_F$. The EDCs at the Fermi momenta for QWS2 (pink and red curves) and QWS1 (orange and green curves) shown in Figs.3d-3e set an upper limit of 3 meV on the gap size for these QWSs. Data taken with other photon energies, e.g. 48 eV and 52 eV, where there are contributions from both the SS and the QWSs, show that the gap size increases consistently with increasing contribution from the SS (see supplementary information). This observation further supports that the large gap does reside on the SS and that the gap on the QWSs is significantly smaller. We note that the gap on the SS is unlikely to be detected directly by STM measurements, since STM lacks the capability to distinguish SS from bulk QWSs. Moreover, STM signals are dominated by QWSs which have negligibly small gap and yet have much higher density of states near the Fermi energy (e.g. see flatter band from QWS1 near E$_F$).

The temperature dependent ARPES data measured with 50 eV photon energy further reveal that the gap on the SS is induced by  superconductivity in Bi2212. Figure 4 shows EDCs at Fermi momenta measured at various temperatures on a sample with 7 meV gap. The gap is observed at least up to 50 K. Above 50 K, the gap becomes much harder to detect and completely vanishes at 100 K. To check reproducibility, we have subsequently cooled down the sample from 110 K directly to 20 K and confirmed that the gap at 20 K was reproducible (bottom curves in Figs.4a-b). Fig.4c shows the extracted gap size as a function of temperature. The similar temperature dependent behavior as the superconductivity transition for Bi2212 clearly suggests that the gap observed on the SS of Bi$_2$Se$_3$ is closely related to superconductivity in Bi2212 through proximity effect.

The coupling between a $d$-wave superconductor and a TI is an intriguing question. The observation of nearly isotropic superconducting gap on the topological SS and much smaller gap on the bulk conduction bands, is a reflection of such nontrivial coupling. If the crystalline group of the Bi$_2$Se$_3$ film were tetragonal as Bi2212, the proximity-induced pairing gap in the Bi$_2$Se$_3$ film would be expected to be nodal $d$-wave in the clean limit. However, since the Bi$_2$Se$_3$ film possesses hexagonal crystal symmetry, the proximity-induced paring gap in the Bi$_2$Se$_3$ is in principle a mixture of $s$-wave and $d$-wave. The $s$-wave component of the induced pairing in the Bi$_2$Se$_3$ is zero when the nodal direction of Bi2212 lies in one of the reflection planes of the Bi$_2$Se$_3$ for the following symmetry reasoning. If this occurs, the hybrid Bi$_2$Se$_3$/Bi2212 system would obey the reflection symmetry along the nodal direction. Under such reflection, the $s$-wave pairing in the Bi$_2$Se$_3$ is even whereas the $d$-wave pairing in Bi2212 is odd.  Fortunately, it is not the case for the Bi$_2$Se$_3$/Bi2212 heterostructure here (see Fig.1b and Fig.1f). Indeed, it turns out that the Bi$_2$Se$_3$ reflection directions are at 15$^\circ$ or 45$^\circ$ from cuprate nodal directions (supplementary information); consequently a sizable $s$-wave component is expected in the induced pairing of the TI. Furthermore, because of the smaller gap and longer coherence length $\xi \approx 2\hbar v_F/(\pi\Delta) \approx 14.4 nm$ in Bi$_2$Se$_3$ compared with cuprates ($\approx  1 nm$), the $d$-wave gap is much more fragile against disorder and the formation of small domains; however, the $s$-wave gap is robust by Anderson's theorem \cite{AT}. Experimentally, we have observed that the pairing gap on the TI's SS is relatively large and nearly isotropic in the entire momentum space (see Fig.2g) although the pairing gap on the QWSs is negligible. A plausible explanation for such distinction is that the TI's SS have a predominant $s$-wave pairing gap which is robust against disorder as well as the formation of domains, whereas the bulk QWSs have predominant $d$-wave pairing gap which is significantly suppressed by disorder. If the gap on the topological SS of Bi$_2$Se$_3$ is confirmed to be $s$-wave, this will open up a large range of experimental opportunities for observing and engineering a single Majorana zero mode. For instance, STM measurements might be able to distinguish this single Majorana zero mode in a vortex core from other low-lying quasi-particle bound states since the energy splitting between them is on the order of $\Delta^2/\epsilon_F\sim 0.5$ meV, which is within the energy resolution of the state-of-art STM.  We note that non-Abelian statistics of the putative Majorana zero mode in a vortex core may not be sufficiently robust due to the QWSs with vanishingly small gap. Tuning the chemial potential away from  QWSs should avoid such complication from QWSs.

\begin{figure*}
\includegraphics[width=14.8 cm] {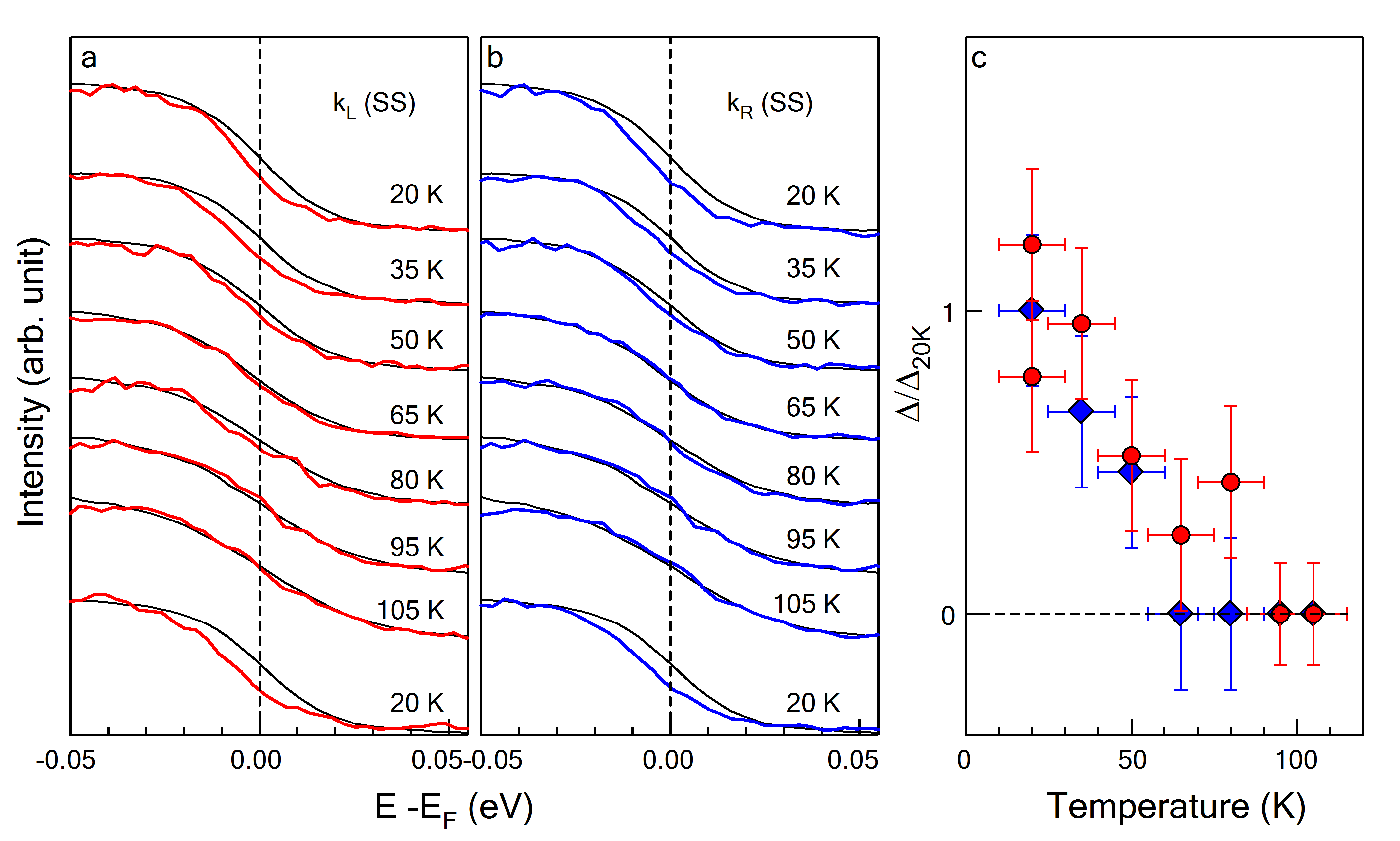}
\label{Fig4}
\caption{{\bf Temperature dependence of the gap on the SS measured with 50 eV photon energy.} (a,b) EDCs at k$_L$ and k$_R$ (labeled in Fig.~2b) measured at different sample temperatures. Black curves were reference spectra for E$_F$. (c) Extracted gap as a function of temperature. The gap values are normalized by the average gap size at 20 K. The gap is defined when the shifted EDC at k$_L$ (k$_R$) shows best overlap with the E$_F$ reference spectrum, and the error bar is defined when the shifted EDCs start to show clear deviation from the E$_F$ reference spectrum. }
\end{figure*}

{\bf Methods}

The MBE growth and {\it in situ} STM experiments were conducted in an MBE-STM combined system with a base pressure of 1.0$\times10^{-10}$ Torr. Single crystal Bi2212 samples with optimum oxygen doping ($T_c$ $\approx$ 91 K) were grown by traveling floating zone \cite{GuGD}.  The Bi2212 samples were cleaved in UHV at room temperature, and annealed at 250 $^{\circ}$C for 3 hours to degas the sample holder to avoid contamination of the film caused by outgassing of the sample holder during the growth process. The atomic structures of Bi atoms on the BiO terminated surface of Bi2212 and the well-known supermodulation in the STM topographic image (Fig.~1b) \cite{DavisSci99} reveal the high quality of the cleaved surface. The characteristic V-shaped gap of 45 meV in the dI/dV spectrum in Fig.~1c confirms that the Bi2212 is still superconducting at near optimum doping and there is only slight change in the oxygen doping after annealing (Fig.S1 in supplementary information).

Bi$_2$Se$_3$ films were then grown on Bi2212 at 190 $^{\circ}$C by co-evaporating Bi (99.9999$\%$) and Se (99.999$\%$) from standard Knudsen cells under Se-rich condition with a growth rate of approximately 0.07 QL per minute \cite{XueBi2Se3}. STM images were taken at 4.2 K with a polycrystalline PtIr tip. The STM image in Fig.~1d shows an atomically flat Bi$_2$Se$_3$ film on Bi2212 with a nominal thickness of 7 quintuple layers (QL). Small regions of 6 QL are also present as indicated by the line profile in Fig.~1e. The step height of 0.95 nm is in agreement with the thickness of 1 QL Bi$_2$Se$_3$ \cite{XuePRL10}. Atomically-resolved STM image (Fig.~1f) reveals a hexagonal lattice structure with a lattice constant of $\approx$ 0.4 nm, consistent with the Se terminated (111) surface of Bi$_2$Se$_3$ \cite{XuePRL10}.

ARPES measurements were taken at BL12.0.1 of the Advanced Light Source at Lawrence Berkeley National Laboratory. Before transferring from the STM chamber to ARPES chamber, the samples were covered with an amorphous Se protection capping layer, which was removed in the APRES chamber by annealing at 200 $^{\circ}$C for 30 minutes until sharp core levels (supplementary information) and clear dispersions were observed. The Fermi energy was calibrated by reference spectra taken from a clean metal (gold or tantalum) surface.

{\bf Acknowledgements}
We thank L. Fu, D.-H. Lee, S. Kivelson and S.-C. Zhang for useful discussions. This work is supported by the National Natural Science Foundation of China (grant No.~11274191 and 11025419) and Ministry of Education of China (20121087903, 20121778394). H.Y. and S.Z. acknowledge the support from the National Thousand Young Talents Program of China.  E.W. acknowledges support from the Advanced Light Source doctoral fellowship program. G.G. and Z.X. are supported by DOE under Contract No. DE-AC02-98CH10886.  J.S. and R.Z. are supported by DOE Center for Emergent Superconductivity. The Advanced Light Source is supported by the Director, Office of Science, Office of Basic Energy Sciences, of the U.S. Department of Energy under Contract No. DE-AC02-05CH11231.

{\bf Author Contributions}
S.Z. and X.C. conceived and designed the experiments. H.D., Z.L., Y-F.L., K.Z., L-G.Z. carried out MBE growth and STM measurements with assistance from S-H.J., L.W., K.H., X.M.,X.C. and Q-K.X. Z.X., J.S., R.Z. and G.G. prepared the bulk Bi2212 samples. E.W., W.Y., A.V.F. and S.Z. performed ARPES measurements and data analysis. S.Z. ,X.C. ,H.Y. and Q-K.X. prepared the manuscript.


\begin {thebibliography} {99}

\bibitem{HasanRMP} Hasan, M.Z. $\&$ Kane, C.L. Colloquium: topological insulators. {\it Rev. Mod. Phys.} {\bf 82}, 3045-3067 (2010).
\bibitem{QiZRMP} Qi, X.-L.$\&$ Zhang, S.C. Topological insulators and superconductors. {\it Rev. Mod. Phys.} {\bf 83}, 1057-1110 (2011).
\bibitem{AliScattering} R. Pedram {\it et al}. Topological surface states protected from backscattering by chiral spin texture. {\it Nature} {\bf 460}, 1106-1109 (2009).
\bibitem{XueScattering} Zhang, T. {\it et al}. Experimental demonstration of topolgoicalsurface states protected by time-reversal symmetry. {\it Phys. Rev. Lett.} {\bf 103}, 266803 (2009).
\bibitem{KanePRB10} J.C.Y. Teo and C.L. Kane. Topological defects and gapless modes n insulators and superconductors. Phys. Rev. B {\bf 82}, 115120 (2010).
\bibitem{FKProximity} Fu, L. $\&$ Kane, C.L. Superconducting proximity effect and Majorana fermions at the surface of a topological insulatr. {\it Phys. Rev. Lett.} {\bf 100}, 096407 (2008).
\bibitem{Kitaev} Kitaev, A. Y. Fault-tolerant quantum computation by anyons, {\it Ann. Phys. (N.Y.)} {\bf 303}, 2-30 (2003).
\bibitem{RMPQCP} Nayak, C. {\it et al}. Non-Abelian anyons and topological quantum computation. {\it Rev. Mod. Phys.} {\bf 80}, 1083-1159 (2008).
\bibitem{MajoranaSci} Mourik, V. {\it et al}. Signatures of Majorana fermions in hybrid superconductor-semiconductor nanowire devices. {\it Science} {\bf 336}, 1003-1007 (2012).
\bibitem{Josephson} Rokhinson, L.P. {\it et al}. The fractional a.c. Josephson effect in a semiconductor-superconductor nanowire as a signature of Majorana fermions. {\it Nature Phys.} {\bf 8}, 795-799 (2012).
\bibitem{XuNanoLett} Deng, M.T. {\it et al}. Anomalous zero-bias conductance peak in a Nb-InSb nanowire-Nb hybrid device. {\it Nano Lett.} {\bf 12}, 6414-6419 (2012).
\bibitem{DasNaturePhys} Das, A. {\it et al}. Zero-bias peaks and splitting in an Al-InAs nanowire topological superconductor as a signature of Majorana fermions. {\it Nature Phys.} {\bf 8}, 887-895 (2012).
\bibitem{FranzReview} Franz, M. Majorana's wires. arXiv:1302.3641 (2013).
\bibitem{QianDBi2Se3} Wang, M. {\it et al}. The coexistence of superconductivity and topological order in the Bi$_2$Se$_3$ thin films. {\it Science} {\bf 336}, 52-55 (2012).
\bibitem{degennes-book} de Gennes, P. G. {\it Superconductivity of Metals and Alloys} (Addison-Wesley, Reading, MA, 1989).
\bibitem{AdvCuprate} Lucignano, P. {\it et al}. Advantages of using high-temperature superconductor heterostructures in the searh for Majorana fermions. {\it Phys. Rev. B} {\bf 86}, 144513 (2012).
\bibitem{Nagaosa} Linder, J. {\it et al}. Unconventional superconductivity on a topological insulator. {\it Phys. Rev. Lett.} {\bf 104}, 067001 (2010).
\bibitem{BurchNatCom} Zareapour, P. {\it et al}. Proximity-induced high temperature superconductivity in the topological insulators Bi$_2$Se$_3$ and Bi$_2$Te$_3$. {\it Nature Comm.} {\bf 3}, 1056 (2012).
\bibitem{DamascelliReview} Damascelli, A. {\it et al}. Angle-resolved photoemission studies of the cuprate superconductors. {\it Rev. Mod. Phys.} {\bf 75}, 473-541 (2003).
\bibitem{HasanBi2Se3} Xia, Y. {\it et al}. Observation of a large-gap topological-insulator class with a single Dirac cone on the surface. {\it Nature Phys.} {\bf 5}, 398-402 (2009).
\bibitem{ChenYL} Chen, Y.L. {\it et al}. Massive Dirac fermions on the surface of a magnetically doped topological insulator. {\it Science} {\bf 329}, 659-662 (2010).
\bibitem{XueBi2Se3} Zhang, Y. {\it et al}. Crossover of the three-dimensional topological insulator Bi$_2$Se$_3$ to the two-dimensional limit. {\it Nature Phys.} {\bf 6}, 584-588 (2010).
\bibitem{PanPRL} Pan, Z.-H. {\it et al}. Electronic structure of the topologicla insulator Bi$_2$Se$_3$ using angle-resolved photoemission spectroscopy: evidence for a nearly fullsurface spin polarization. {\it Phys. Rev. Lett.} {\bf 106}, 257004 (2011).
\bibitem{ShenLEG} Shen, Z.X. {\it et al}. Anomalously large gap anisotropy in the A-B plane of Bi$_2$Sr$_2$CaCu$_2$O$_{8+\delta}$. {\it Phys. Rev. Lett.} {\bf 70}, 1553 (1993).
\bibitem{AT} Anderson, P.W. Theory of dirty superconductors. {\it J. Phys. Chem. Solids} {\bf 11}, 26-30 (1959).
\bibitem{GuGD} Gu, G.D. {\it et al}. Growth and superconductivity of Bi$_{2.1}$Sr$_{1.9}$Ca$_{1.0}$(Cu$_{1-y}$Fe$_y$)$_2$O$_x$ single crystal. {\it J. Cryst. Growth} {\bf 137}, 472-478 (1994).
\bibitem{DavisSci99} Hudson, E.W. {\it et al}. Atomic-scale quasi-particle scattering resonances in Bi$_2$Sr$_2$CaCu$_2$O$_{8+\delta}$. {\it Science} {\bf 285}, 88-91 (1999).
\bibitem{XuePRL10} Cheng, P. {\it et al}. Landau quantization of topological surface states in Bi$_2$Se$_3$. {\it Phys. Rev. Lett.} {\bf 105}, 076801 (2010).


\end {thebibliography}
\end{document}